\documentclass[aps]{revtex4}
\begin{document}
\textheight 8.8in
\textwidth 6.5in
\topmargin -.25in
\oddsidemargin -.25in
\evensidemargin 0in
\baselineskip 14pt
\newcommand{\lsim}{\,\lower2truept
\hbox{${<\atop\hbox{\raise4truept\hbox{$\sim$}}}$}\,}
\newcommand{\gsim}{\,\lower2truept
\hbox{${>\atop\hbox{\raise4truept\hbox{$\sim$}}}$}\,}
\title{{\large {\bf Non-linear perturbations in scalar-tensor cosmologies}}}
\author{F. Perrotta$^{1,2,3}$, S. Matarrese$^{4,5}$, M. Pietroni$^{5}$, 
C. Schimd$^{6,7}$} 
\affiliation{
$^1$ SISSA/ISAS, Via Beirut 4, 34014 Trieste, Italy; \\
$^2$ INFN-Sezione di Trieste, via Valerio 2, 34127 Trieste, Italy;  \\
$^3$ Lawrence Berkeley National Laboratory, 1 Cyclotron road, Berkeley 
94720, CA; \\
$^4$ Dipartimento di Fisica `G. Galilei', Universita di Padova, via 
Marzolo 8, 35131 Padova, Italy; \\
$^5$ INFN-Sezione di Padova, via Marzolo 8,  35131 Padova, Italy; \\
$^6$ Dipartimento di Fisica `M. Melloni', Universita' di Parma; \\
$^7$ INFN-Gruppo collegato di Parma, Parco Area delle Scienze 7/A, 43100 
Parma, Italy}

\begin{abstract}  
Can local fluctuations of a ``Quintessence'' scalar field play a dynamical 
role in the gravitational clustering and cosmic structure formation 
process? We address this question in the general framework of scalar-tensor 
theories of gravity. 
Non-linear energy density perturbations, both in the scalar 
field and matter component, and linear metric perturbations are 
accounted for in the perturbed Einstein's equations. 
We derive the Newtonian limit of the relevant equations for 
clustering in scalar-tensor cosmologies. 
We then specialize to non-linear  perturbations of the ``Extended 
Quintessence'' model of Dark Energy; 
in such a model, a non-minimally coupled scalar field is thought to be 
responsible for driving the present accelerated phase of the 
Universe expansion. The interplay between Dark Energy and Dark Matter is 
displayed in the equations governing the growth of structure in the Universe. 
\end{abstract}

\maketitle

\section{Introduction} 
\label{introduction}
Our traditional picture of the Universe has been definitely upset when, in 
1998, astronomers found that distant type 1a Supernovae were dimmer than 
expected in a decelerating Universe \cite{Perl}, \cite{Riess}. 
This early evidence has been confirmed by the subsequent studies, which
combined the most recent Cosmic Microwave Background (CMB) data from WMAP 
(see \cite{Spergel}, \cite{Bennett} and references therein) and 
Large Scale Structure (LSS) data \cite{Dodelson}, \cite{Percival,Verde}, 
together with measurements of the Hubble constant \cite{Freedman}; 
the fact that the Universe turns out to be geometrically close to flat,  
together with the estimates of its matter content, has called for deep 
changes of the old ``standard'' scenario of a matter dominated Universe.  
According to these observations, almost $70 \%$ of the total energy of 
the Universe resides in a ``Dark Energy'' component, which is plausibly 
acting as a repulsive force driving the cosmic acceleration. 
Although the Cosmological Constant has been historically proposed as a 
candidate for the cosmic acceleration, the theoretical difficulties in
justifying its exceedingly small value motivated the search for alternative 
theories. \\
There are currently two different general approaches to Dark Energy modeling; 
one class of models is based on the introduction of a new cosmological 
component having negative equation of state, often described 
through the dynamics of a self-interacting, minimally coupled scalar 
field \cite{RP}-\cite{WS} . 
The other class 
of models  proposes modifications to the gravity itself, introducing a 
non-minimally coupled scalar field (scalar-tensor theories)  or changing 
the function of the Ricci scalar appearing in the  Gravitational sector 
of the Lagrangian of the theory \cite{Chiba}-\cite{CMM}.  
In both pictures, an open important issue is whether modifications of 
gravity, or small-scale perturbations of a Quintessence scalar field, 
could lead to significant effects on the formation of
structures, such as galaxies and clusters (see \cite{Amendola2}, 
\cite{PB},\cite{Wett2}-\cite{FP}). \\
A powerful tool for the investigation of the Dark Energy main properties 
and parameters has been the linear perturbation theory, which allowed 
to make accurate predictions on many cosmological observables, such as 
the CMB spectrum of anisotropies (see, e.g., \cite{const1}-\cite{const5}); 
in this paper, we deal with the behavior of perturbations in the 
non-linear regime. We focus on scalar-tensor theories, where a scalar field 
non-minimally coupled to the Ricci scalar is proposed as a candidate for 
the Dark Energy component; in principle, the ``Extended Quintessence'' 
field \cite{PBM} allows for non-vanishing small-scale perturbations, 
which have been analyzed in the 
linear regime in \cite{PBM}, \cite{BMP}, \cite{PB}, \cite{Boiss}. In 
particular, in \cite{Boiss} and \cite{PB} it was shown that, while 
perturbations in a minimally-coupled scalar field behave as radiation on 
sub-horizon scales \cite{hu}, 
so that the field rapidly becomes a smooth component on such scales, 
perturbations in a non-minimally coupled scalar field can be dragged by 
perturbations in the matter component, thanks to the coupling of the 
scalar field to the Ricci scalar, and become non-linear. 
In practice, the mechanism which would damp out the field perturbations 
is here counterbalanced by the presence of a source term, which is 
directly related to the gravitational potentials and, ultimately, to the 
matter perturbations: we expect this dragging to be 
able to affect gravitational clustering down to galactic scales. \\
We extend here the analysis performed in  \cite{PB} for the 
linear perturbations in scalar-tensor theories, obtaining the non-linear 
Newtonian limit of Einstein's equations for this class of models. 
These will be the equations to be eventually inserted in a modified N-body 
code, in order to simulate the behavior of collapsing matter under this 
modified property of gravity. Some numerical simulations, involving Dark 
Energy in several context, have been performed in Refs.  
\cite{MCBW}-\cite{DBPBMMT}.  \\
Finally, we focus on the Extended Quintessence model, specializing the 
Poisson equation for the gravitational potential in order to evaluate 
modifications with respect to the ``standard'' theory. \\
The plan of the paper is as follows: in section \ref{general}, we write 
the perturbed Einstein's equations and the equation for scalar 
field perturbations in non-linear regime, assuming that the metric 
perturbations are linear, while those of matter and scalar field are not.   
In sect. \ref{Newton} we discuss the approximations giving the 
Newtonian limit of these equations, and in  sect. \ref{specmodel} we 
specify the  general equations for scalar-tensor theories to the 
``Extended Quintessence'' case. Finally, in sect. \ref{concl}
we draw our conclusions.  

\section{Perturbed Einstein equations in Extended Quintessence }
\label{general} 
Since we are interested in the problem of how Dark Energy perturbations 
could affect the structure formation process, excluding the limit of 
very strong gravitational fields (typical, for example, of black holes), 
we will consider linear perturbations of the metric tensor, while keeping 
non-linearity in the matter and scalar field  
perturbations \cite{MPS}. This also means that we are restricting our
study to non-relativistic Dark Matter, whose typical velocity is much smaller
than the speed of light  even in the presence of highly non-linear matter 
overdensities. \\ 
Metric perturbations can be decomposed into scalar modes, as well 
as by vector and tensor ones; by taking the spatial covariant divergence of 
the ($0$-$i$) Einstein equations and the trace of the ($i$-$j$) equations, 
we will single out scalar modes, so that, without loss of generality, the 
line-element can be written, in conformal Newtonian-gauge formalism, in terms 
of only two gravitational potentials $\Phi$ and $\Psi$ as  
\begin{equation}
ds^2=a^2(\eta)[-(1+2 \Phi)d\eta^2+(1-2\Psi)\delta^i_idx^idx_j]=
g_{\mu \nu} dx^{\mu}dx^{\nu} \ \ ,
\end{equation}
where $\eta$ denotes conformal time; note that we adopt the signature 
(--+++). The Lagrangian for scalar-tensor theories reads   
\begin{equation}
\label{Lagr}
{\cal{L}}={1 \over 2 \kappa} f(\phi,R)-{1\over 2} 
\omega(\phi)\phi_{,c}\phi^{,c}-V(\phi)+{\cal{L}}_{matter}
\end{equation}
where $\kappa \equiv 8 \pi G^*$ and $G^*$ is the bare gravitational 
constant. \\
We will consider a Universe filled by dark matter and a non-minimally 
coupled scalar field; the matter component will be described in terms a 
discrete set of particles with constant mass $m$ and coordinates 
${\bf x}_a(\eta)$ $(a=1,2,...)$. Following \cite{Weinberg}, if we denote 
the matter four-velocity by $u^{\mu}\equiv dx^{\mu}/ d x^0 $ ($x^0 
\equiv \eta$), the matter stress-energy tensor reads 
\begin{equation}
\label{tmatter}
T^{\mu}_{\nu} \sim a^{-2} \rho_m u^{\mu} u_{\nu}
\end{equation} 
where we neglected corrections ${\cal{O}}|{\bf u}|^2$, and we defined 
\begin{equation} 
\rho_m = m a^{-3} \sum_{a} \delta^{(3)}({\bf x}-{\bf x}_a)  \ \ \ .
\end{equation} 
The scalar field stress-energy tensor reads 
\begin{equation}
\label{stressenergy}
T^{\alpha}_{\beta }=\omega\left[ g^{\alpha \nu}\phi_{,\nu}\phi_{, 
\beta}-{1 \over 2}g^{\alpha \nu}
g_{\nu \beta}\phi_{, \sigma}\phi^{,\sigma}\right]-g^{\alpha \nu}g_{\nu 
\beta}V+ {1\over 2} g^{\alpha \nu}g_{\nu \beta}\left({f \over \kappa}-RF 
\right)+g^{\alpha \nu}F_{, \nu ; \beta}-g^{\alpha \nu}g_{\nu \beta}
{F^{,\sigma}}_{; \sigma}+
\left({1 \over \kappa}-F\right) G^{\alpha}_{\beta}
\end{equation}
where, $R$  is the Ricci scalar and
$ F \equiv {1 \over \kappa } {\partial f \over \partial R}$. 
Restricting ourselves to the simplest case of $f(\phi, R)$  linear in 
$R$, (as is the case of the ``Extended  Quintessence'' model), it turns out 
that  $FR= {f(\phi,R) \over \kappa}$.  
We also assume that  $\omega=const=1$, so that the tensor (\ref{stressenergy}) 
becomes
\begin{equation}
\label{tsf}
T_{\alpha \beta}=\phi_{,\alpha} \phi_{,\beta}-{1 \over 2} 
g_{\alpha \beta }\phi_{, \sigma}\phi^{, \sigma}-Vg_{\alpha \beta}
+F_{, \alpha; \beta}-g_{\alpha \beta }{F^{, \sigma}}_{; \sigma}+\left( {1 
\over \kappa}-F\right) G_{\alpha \beta }
\end{equation}
Both the scalar field and the function $F(\phi({\bf x},\eta))$
 can be written as the superposition of a 
background component, only dependent on time, plus a generally 
non-linear perturbation, which depends both on the spatial coordinates 
and on time:  
\begin{equation}
\nonumber
\phi({\bf x},\eta) \equiv \phi_{\rm b}(\eta) + 
\delta \phi({\bf x},\eta) \ \ \ \ \ 
F(\phi({\bf x},\eta)) \equiv F_{\rm b}(\eta) + \delta F({\bf x},\eta) \ \ ,
\nonumber
\end{equation}
The evolution of the background quantities is fully determined by the 
Friedmann equations and the unperturbed Klein-Gordon equation, which 
read, respectively, 
\begin{equation} 
\label{Friedmann1}
{\cal H}^2 = {1 \over 3 F_{\rm b}} 
\left( a^2 \langle\rho_m\rangle+ {1\over 2 } 
\dot{\phi}_{\rm b}^2
+a^2 \langle V \rangle-3 {\cal H} \dot{F}_{\rm b} \right)  \ \ \ ;
\end{equation}
\begin{equation} 
\label{Friedmann2}
\dot{{\cal H}} = -{1 \over 6F_{\rm b}} \left( a^2  \langle\rho_m \rangle
+ 2 \dot{\phi}_{\rm b}^2-2 a^2 \langle V \rangle
+3 \ddot{F}_{\rm b} \right) \ \ \ ,  
\end{equation}
where  $\langle V \rangle \equiv V(\phi_{\rm b}(\eta))$ and  
$\langle \rho_m \rangle$ is the background value of the matter energy-density. 
Finally, the background Klein-Gordon equation in scalar-tensor 
theories reads 
\begin{equation}
\label{KGbackground}
\ddot{\phi_{\rm b}}+2{\cal H}\dot{ \phi_{\rm b}}-{1 \over 2} a^2 R_{\rm b} 
F_{\rm b}' + a^2 V'(\phi_{\rm b})=0 \ \ , 
\end{equation}
where a prime denotes differentiation with respect to $\phi$. \\
In our approach to the perturbed Einstein's equations, 
the Ricci and Einstein's tensors will be linearly perturbed, while the 
perturbations in the components of the stress-energy  tensors 
(\ref{tmatter}) and (\ref{tsf}) are obtained by subtracting their 
background (mean) values from the fully non-linear ones: 
\begin{equation}
\delta G^{\mu}_{\nu}= \kappa \left( T^{\mu}_{\nu}- \langle 
T^{\mu}_{\nu} \rangle \right) 
\ \ \ . 
\end{equation}
As for the ($0$-$0$) component (``energy constraint''), we get the following 
perturbation equation: 
\begin{eqnarray}
\nonumber
 \nabla ^2  \Phi &+& 3 {\cal H}(\dot\Phi + 
\dot\Psi)
+3 \ddot\Psi= 8 \pi G^* \left[
{ 1\over 2} \left( \sum_{a} a^{-1}m_a \delta^{(3)}_{(a)}-a^2 \langle\rho_m 
\rangle 
\right)+\sum_{a} a^{-1}m_a \delta^{(3)}_{(a)}\Phi
+ \right. \\ 
\nonumber
&+& (\delta \dot{\phi})^2+2 \dot{\phi}_{\rm b} \delta \dot{\phi}-a^2  
(V- \langle V \rangle)-2a^2\Phi V + \\ 
\nonumber
&+& {3 \over 2 } \delta \ddot{F}-{3 \over 2 } \dot{ F} (\dot \Phi 
+ \dot\Psi)+{1\over 2 } \nabla\delta F \cdot 
 (\nabla\Psi-3\nabla \Phi)
-{1 \over 2}\nabla^2 \delta F (1+2\Phi+2\Psi)+ \\
\label{00}
&+& \left. 3 \dot{\cal H}\delta F+
\left( {1 \over \kappa}- F\right)
(3 {\cal H} \dot\Psi +3 \ddot\Psi
+3 {\cal H} \dot\Phi + \nabla^2\Phi) \right] \ \  , 
\end{eqnarray}
where $\delta^{(3)}_{(a)} \equiv \delta^{(3)}({\bf{x}}-{\bf{x}}_a)$. \\
Similarly, the equation for the divergence of the ($0$-$i$)
or ``momentum constraint'' equation (herefrom latin indices will be used 
to label spatial coordinates) reads 
\begin{eqnarray}
\nonumber
&{\cal H}&{\nabla}^2 \Phi+{\nabla}^2\dot\Psi = 4 \pi G^*
\left[-\sum_{a}m_a a^{-1}
\left(\delta^{(3)}_{(a)} u^i_{(a)} \right)_{,i} 
+ 2\dot{\phi} \nabla\Psi\cdot \nabla \delta \phi
+(1+2\Psi) \dot{\phi} \nabla^2 \delta \phi+
(1+2\Psi) \nabla \delta \phi \cdot \nabla\delta \dot\phi
+ \right. \\  
\nonumber
&+& \nabla\delta F \cdot \left( 2 {\cal H}\nabla \Psi
-\nabla \dot\Psi \right)+
\nabla^2 \delta F \left(-{\cal H}-2{\cal H}\Psi +
\dot\Psi\right) + \nabla \delta \dot{F} 
\cdot\left(2\nabla \Psi - \nabla \Phi\right)+
\nabla^2 \delta \dot{F} (1+ 2 \Psi)-\dot{F} \nabla^2\Phi-\\
\label{0i}
&-& \left. 2 \nabla\delta 
F\cdot({\cal H}\nabla\Phi + \nabla\dot\Psi)
+2\left({1\over \kappa}-F \right)\left( {\cal H} \nabla^2 
\Phi+\nabla^2\dot\Psi \right) \right]
\end{eqnarray}
Finally, the trace of the ($i$-$j$) component is: 
\begin{eqnarray}
\nonumber
&-& 2\dot{\cal H}\Phi-{\cal H}(\dot\Phi+
5\dot\Psi)-4{\cal H}^2\Phi-\ddot\Psi
+{4 \over 3}\nabla^2\Psi - {1 \over 3}{\nabla}^2 \Phi = \\
\nonumber
&=& 8 \pi G^* \left[  
{1 \over 2}\left( \sum_{a} m a^{-1} \delta^{(3)}_{(a)}-a^2 
\langle \rho_m \rangle\right)+
{1 \over 3}(1+2 \Psi) |\nabla\delta \phi|^2+a^2(V- \langle V \rangle)+  
\right. \\
\nonumber
&+& \Phi \ddot{F}|_{\rm b}-{1 \over 2}(1-2 \Phi)\delta \ddot{F}-
2 {\cal H} 
\delta \dot {F}+\dot {F} (4{\cal H}\Phi + {1 \over 
2}\dot\Phi + {5 \over 2}\dot\Psi)+\nabla\delta F 
\cdot(-{5\over 6}\nabla\Psi + {1 \over 2}\nabla\Phi)
+{5\over 6}\nabla^2\delta F(1+2 \Psi)+ \\
\label{ii} 
&+& \left. \left({1 \over \kappa}-F\right)
\left(-2 \Phi(\dot{\cal H}+2{\cal H}^2)-{\cal H}\dot\Phi
-{1\over 3}\nabla^2 \Phi-5{\cal H}\dot \Psi -\ddot\Psi +
{4 \over 3}\nabla^2\Psi \right) -\delta 
F(\dot{\cal H}+2{\cal H}^2) \right]
\end{eqnarray}
The perturbed Klein-Gordon equation gives    
\begin{eqnarray}
\nonumber
&\delta \ddot{\phi}&+2{\cal H}\delta{\dot{ 
\phi}}-(\dot\Phi + 3
\dot\Psi)(\dot \phi_{\rm b} +\delta\dot{\phi})-\nabla(\Phi
- \Psi)\cdot \nabla \delta \phi
-(1+2 \Phi +2 \Psi ) \nabla^2 \delta \phi 
+a^2[V'(\phi({\bf x},\eta))-V'(\phi_{\rm b})]+ \\ 
&2&a^2 \Phi V'(\phi({\bf x},\eta))-
{a^2 \over 2} \delta F' R_{\rm b}+(F'_{\rm b}+\delta F')(3 {\cal H}\dot \Phi 
+\nabla^2 \Phi + 3\ddot\Psi + 9 {\cal H}\dot\Psi - 2\nabla^2 
\Psi)=0
\end{eqnarray}
where $R_{\rm b}=6 a^{-2}(\dot{\cal H}+{\cal H}^2)$ is the Ricci scalar in the 
unperturbed metric. 
Finally, the equation describing the motion of Dark Matter particles, in the 
weak-field limit, is given by 
\begin{equation} 
\label{motion}
\ddot{x}^i_a  + \dot{x}^i_a\left[ 
{\cal H}-\dot\Phi-2\dot\Psi
\right]=-\partial^i \Phi \ \ \ . 
\end{equation}
The system (\ref{00})-(\ref{motion}) is a closed system of equations in the 
variables $\Phi, \Psi, \phi, x^i_a$'s, which is redundant in the 
number of equations, as a consequence of the underlying 
gauge-invariance of Einstein's theory; as we will see, it will be more 
convenient to disregard 
the momentum constraint, working with the remaining equations. \\
All these equations apply to cosmological scales, under the assumption 
of linear perturbations of the gravitational potentials. In the next 
section, we will determine the system adequate to determine the matter 
particle motion on the scales relevant for structure formation, 
extending the standard Newtonian approximation to the case of 
scalar-tensor theories.  

\section{The extended Newtonian approximation}
\label{Newton}
There are many considerations that can help to simplify the equations 
(\ref{00})-(\ref{motion}). First of all, note that, in the weak-field 
limit, $|\Phi|$ and $|\Psi|$ are of the order of $|u^i|^2 \ll 
1$, $|u^i|$ being the typical velocity of non-relativistic matter; 
furthermore, since the characteristic evolution time of $\Psi$ and 
$\Psi$ is $\tau_{dyn} \sim \tau \equiv {\cal H}^{-1}$, we can 
neglect $\dot\Phi, \dot\Psi$
with respect to ${\cal H}$ in the energy constraint equation, in the 
perturbed Klein-Gordon and in the equation of motion. \\
Focusing on the perturbation growth on scales well below the Hubble radius, 
we can apply further approximations to the system. 
On those scales, $\ddot\Phi,{\cal H}\dot\Phi \sim 
{\cal H}^2 \Phi 
\ll \nabla^2\Phi$ and 
$\ddot\Psi,{\cal H}\dot\Psi \sim {\cal H}^2 \Psi 
\ll \nabla^2 \Psi$ and we can neglect such terms in the energy 
constraint, Einstein spatial trace and Klein-Gordon equations. We also 
assume that $\dot{\cal H} \sim {\cal H}^2$, so the last term on the 
right-hand side of the trace equation is negligible, on sub-horizon 
scales, with respect to the term containing $\nabla ^2 \delta F$. Equations 
(\ref{00})-(\ref{motion}) reduce therefore to the 
following system:  \\
\begin{eqnarray}
\nonumber
&\nabla ^2&  \Phi= 8 \pi G^* \left[
{ 1\over 2} \left( \sum_{a} a^{-1}m_a \delta^{(3)}_{(a)}-a^2 
\langle \rho_m \rangle \right)+
 (\delta {\dot \phi})^2+2 \dot{\phi_{\rm b}} \delta \dot{\phi}-a^2  
(V- \langle V \rangle)+ \right. \\ 
\label{00e}
&+& {3 \over 2 } \delta \ddot{F}-{3 \over 2 } \dot{ F} (\dot \Phi 
+ \dot\Psi)-{1 \over 2}{\nabla}^2 \delta F 
+ \left. 3 \dot{\cal H} \delta F+
\left( {1 \over \kappa}- F\right)
{\nabla}^2 \Phi \right] \ \ \ ;
\end{eqnarray}
\begin{eqnarray}
\nonumber
&{\cal H}&\nabla^2\Phi+ \nabla^2\dot\Psi = 4 \pi G^*
\left[-\sum_a m_a a^{-1} \left( \delta^{(3)}_{(a)}u^i_{(a)}\right)_{,i} + 
\dot{\phi} \nabla^2 \delta \phi+
\nabla \delta \phi \cdot \nabla\delta \dot{\phi} - \right. \\
\label{mome}
&-& {\cal H}\nabla^2\delta F+
\nabla^2 \delta \dot {F} -\dot{F} \nabla^2\Phi
+  \left. 
2\left({1\over \kappa}-F \right)\left( {\cal H} \nabla^2 
\Phi+{\nabla}^2\dot\Psi \right) \right] \ \ \ ;
\end{eqnarray}
\begin{eqnarray}
\nonumber
{4 \over 3} & {\nabla}^2 & \Psi - {1 \over 3}\nabla^2\Phi = 
 8 \pi G^* \left[  {1 \over 2}\left( \sum_a m a^{-1} \delta^{(3)}_{(a)}-a^2 
\langle\rho_m \rangle\right)+
{1 \over 3} |\nabla\delta \phi|^2+a^2(V- \langle V \rangle)+ \right. \\
\label{tracee}
&+& \Phi\ddot{F}_{\rm b} - {1 \over 2}\delta \ddot{F}-2 
{\cal H} 
\delta \dot {F}+\dot{F}(4{\cal H}\Phi + {1 \over 
2}\dot\Phi + {5 \over 2} \dot\Psi)
+{5\over 6}\nabla^2\delta F+ 
 \left. \left({1 \over \kappa}-F\right)
{ \nabla^2 \over 3}  \left(-\Phi + 4 \Psi \right) \right] \ \ ;
\end{eqnarray}
\begin{eqnarray}
\nonumber
\delta \ddot{\phi} &+& 2{\cal H}\delta \dot{ 
\phi}-(\dot\Phi+3
\dot\Psi) \dot \phi_{\rm b} - \nabla^2 \delta \phi 
+a^2[V'(\phi({\bf x},\eta))-V'(\phi_{\rm b})]+ \\ 
\label{klein}
&-&
{a^2 \over 2} \delta F' R_{\rm b}+(F'_{\rm b}+\delta F') 
(\nabla^2\Phi-2\nabla^2  \Psi)=0 \ \ \ ;
\end{eqnarray}
\begin{equation}
\label{finmot}
\ddot{x}^i_a + {\cal H}\dot{x}^i_a =-\partial^i \Phi
\end{equation}
In order to further simplify the equations, we have to estimate the 
characteristic evolution time for scalar field perturbations. We will 
follow the approach of \cite{MPS}, 
considering  the scalar field perturbation as the sum of a 
``relativistic'' perturbation $\delta \phi_{\rm R}$ and a ``non-relativistic'' 
one,  $\delta \phi_{\rm NR}$, 
\begin{equation}
\label{defin}
\delta \phi ({\bf x}, \eta) \equiv  \delta \phi_{\rm R} ({\bf x}, \eta) + 
\delta \phi_{\rm NR}({\bf x}, \eta)
\end{equation}
where $\delta \phi_{\rm NR}$ is defined as the 
solution of the  perturbed Klein-Gordon equation in the $c 
\rightarrow \infty$ limit. By definition, the time derivative of  $\delta 
\phi_{\rm NR}$  is negligible with respect to that of $\delta \phi_{\rm R}$, 
which, on the contrary, has a wave-like behavior. The characteristic 
(conformal) time for the variation of $\delta \phi_{\rm NR}$, $\tau_{\rm NR}$, 
is much larger than the characteristic time for the variation of $\delta 
\phi_{\rm R}$, $\tau_{\rm R}$: on scales much smaller 
than the horizon, the time-variation of the non-relativistic component of 
scalar field perturbations is negligible with respect to the spatial 
gradient, which is equivalent to the limit 
$c \rightarrow \infty$. \\
The Klein-Gordon equation for $\delta \phi_{\rm NR}$ reduces to 
\begin{equation}
\label{kleinNR}
- \nabla^2 \delta \phi_{\rm NR} 
+a^2 \delta V'(\delta \phi_{\rm NR}({\bf x}, \eta))
+(F'_{\rm b}+\delta F'_{\rm NR})\nabla^2( \Phi-2\Psi)-{a^2 \over 2} 
\delta F'_{\rm NR}R_{\rm b}=0
\end{equation}
where $\delta V'(\delta \phi_{\rm NR}({\bf x}, \eta)) \equiv 
V'(\phi_{\rm b}(\eta)+\delta \phi_{\rm NR}({\bf x}, \eta)) - 
V'(\phi_{\rm b}(\eta))$, and 
$\delta F_{\rm NR} \equiv F(\phi_{\rm b} + 
\delta \phi_{\rm NR})-F(\phi_{\rm b}) $.  \\
Inserting the definition (\ref{defin}) into eq. (\ref{klein}), and using 
(\ref{kleinNR}), we obtain the equation for $\delta \phi_{\rm R}$: 
\begin{eqnarray}
\nonumber
& & \delta  \ddot{\phi}_{\rm R}+ 2 {\cal H} \delta \dot{\phi}_{\rm R}
-\nabla^2 \delta {\phi}_{\rm R}+a^2[V'(\phi_{\rm b} + \delta \phi_{\rm R}+
\delta {\phi}_{\rm NR})- V'(\phi_{\rm b}+ \delta \phi_{\rm NR})]
- {a^2 \over 2 } [F'(\phi_{\rm b}+\delta \phi_{\rm NR}+ 
\delta \phi_{\rm R})- \\ 
&-& F'(\phi_{\rm b}+\delta \phi_{\rm NR})]R_{\rm b} + 
 [F'(\phi_{\rm b} + \delta \phi_{\rm NR}+\delta \phi_{\rm R}) - 
F'(\phi_{\rm b} + \delta \phi_{\rm NR})] \nabla^2(\Phi - 2 \Psi) =0 \ \ \ . 
\label{kgr}
\end{eqnarray}
In the last equation we neglected all terms of order ${\cal H}^2$:
since $\delta \dot{\phi}_{\rm NR}/ \delta {\phi}_{\rm NR} \sim 
\tau_{\rm NR}^{-1}$, we 
have
\begin{equation}
\delta \ddot{\phi}_{\rm NR}/ \delta {\phi}_{\rm NR}  \sim \tau_{\rm NR}^{-2} 
\lsim 
{\cal H}^2; 
\ \ \ \
2 {\cal H} \delta \dot{\phi}_{\rm NR}/ \delta\phi_{\rm NR} \sim 
\tau_{\rm NR}^{-1} 
\tau^{-1} \lsim {\cal H}^2; 
\ \ \ \ 
(\dot\Phi+3 \dot\Psi) \dot{\phi}_{\rm b} \sim {\cal H}^2 .
\end{equation}
We assume that, on the scales characteristic for structure formation, 
$\delta {\phi}_{\rm R}$ is a small 
perturbation (this assumption will be verified {\it a posteriori}); in 
such a case, $V'(\phi_{\rm b} + \delta {\phi}_{\rm R}+\delta {\phi}_{\rm NR})- 
V'(\phi_{\rm b} + \delta \phi_{\rm NR}) \sim  
V''(\phi_{\rm b} + \delta \phi_{\rm NR}) 
\delta \phi_{\rm R}$ , and $F'(\phi_{\rm b} + 
\delta \phi_{\rm R}+\delta {\phi}_{\rm NR})-
F'(\phi_{\rm b} + \delta \phi_{\rm NR}) \sim  F''(\phi_{\rm b} + 
\delta \phi_{\rm NR}) 
\delta \phi_{\rm R}$. Substituting in eq. (\ref{kgr}), there will be
two terms containing  $ F''(\phi_{\rm b} + \delta \phi_{\rm NR}) \delta
\phi_{\rm R}$; they will produce, respectively,
a term of order  $\delta \phi_{\rm R} k^2 (\Phi - 2\Psi) $ and a 
term of order $\delta \phi_{\rm R} {\cal H}^2 $, which are both
negligible with respect to $\nabla ^2 \delta \phi_{\rm R} $. Therefore, 
eq. (\ref{kgr}) becomes 
\begin{equation}
\label{kgrel}
\delta \ddot{\phi}_{\rm R}+ 2 {\cal H} \delta \dot{\phi}_{\rm R}
-\nabla^2 \delta {\phi}_{\rm R}+a^2 V''(\phi_{\rm b} + 
\delta \phi_{\rm NR}) \delta \phi_{\rm R} =0 \;. 
\end{equation}
This equation describes a (quasi-massless) plane-wave, whose amplitude is 
damped by the cosmic expansion, so that in a timescale $\tau \sim 
{\cal H}^{-1}$ the relativistic perturbation $\delta \phi_{\rm R}$ 
vanishes on scales smaller that the horizon, $\delta \phi_{\rm R} \rightarrow 
0$. The mass associated to this wave is $a^2 V''(\phi_{\rm b}+\delta 
\phi_{\rm NR}) \sim {\cal H}^2$, negligible on sub-horizon scales: as the 
perturbation  $\delta\phi_{\rm R}$ enters the horizon, it will behave as 
radiation.    
We have thus verified {\it a posteriori} the validity of our 
assumption on the smallness of $\delta {\phi}_{\rm R}$, which will be 
neglected hereafter. From now on, $\delta \phi$ has to be understood as 
$\delta  \phi_{\rm NR}$, the non-relativistic component of the scalar field 
perturbation, and we will omit the subscript $_{\rm NR}$; this will be, in 
general, a non-linear perturbation. 

In the Newtonian limit we can neglect time derivatives  in 
the trace equation and in the energy constraint. Indeed, the scales which we 
are considering are well below the horizon, so the Laplacian 
terms will dominate over terms of order ${\cal H}^2$. Note that 
$\dot{\phi}_{\rm b}^2$ also is of order ${\cal H}^2$, as a consequence of 
the Friedmann equations and of the smallness of the coupling at low 
redshifts; we will neglect $\dot{\phi}$ in this limit. \\ 
The energy constraint reads: 
\begin{equation}
\label{00f}
F \nabla ^2  \Phi =  \left[
{ 1\over 2} \left( \sum_{a} a^{-1}m_a \delta^{(3)}_{(a)}-a^2 
\langle \rho_m \rangle 
\right)
-a^2 (V- \langle V \rangle)-{1 \over 2}{\nabla}^2 \delta F 
\right]
\end{equation}
and the trace equation reads
\begin{equation}
\label{tracef}
{F \over 3} ( 4 \nabla^2\Psi-\nabla^2 \Phi)= 
 \left[  {1 \over 2}\left( \sum_a m a^{-1} \delta^{(3)}_{(a)}-a^2 
\langle \rho_m \rangle\right)+
{1 \over 3} |\nabla\delta \phi|^2+a^2(V- \langle V \rangle)
+{5\over 6}{\nabla}^2\delta F \right]
\end{equation}
In the equations above, $\delta \phi \equiv \delta \phi_{\rm NR} $. 
Note that, as anticipated in the previous section, the system 
(\ref{00e})-(\ref{tracee}) provides a redundant 
set of equations for the gravitational potentials $\Phi$ and 
$\Psi$: the ($0$-$0$) equation and the trace equation fully specify 
their evolution, so we 
can get rid of the spatial covariant derivative of the ($0$-$i$) Einstein's 
equations. Using eq. (\ref{00f}) and (\ref{tracef}), together with eq. 
(\ref{kleinNR}), we solve for $\Phi, \Psi, {\delta \phi}$:
\begin{equation}
\label{finalPsi}
\nabla ^2 \Psi = { 1 \over 2F } \left[ {a^2 \delta \rho_m } 
 +  {a^2 \delta V }  + \nabla^2 \delta F + 
{1\over 2} |\nabla  \delta \phi |^2  \right] \ \ ;
\end{equation}
\begin{equation}
\label{finalP}
\nabla ^2 \Phi= { 1 \over 2F } \left[ {a^2 \delta \rho_m }
 -2 {a^2 \delta V }  - \nabla^2 \delta F  \right] \ \ ;
\end{equation} 
where $\delta V \equiv V(\phi_{\rm b} + \delta \phi) -V(\phi_{\rm b})$ and 
$a^2 \delta\rho_m  \equiv \sum_{a} a^{-1}m_a 
\delta^{(3)}_{(a)}-a^2 \langle\rho_m\rangle$.
The perturbed Klein-Gordon equation becomes 
\begin{equation} 
\label{finalKG}
\nabla^2 \delta \phi = a^2 \delta V'-{F'_{\rm b}+\delta F' \over 2 F}
\left[ {a^2 \delta \rho_m } 
 +4  {a^2 \delta V }  + 3 \nabla^2 \delta F +
 |\nabla \delta \phi |^2 \right]
-{a^2 \over 2} \delta F' R_{\rm b} 
\end{equation}

Once the function F is specified together with the 
scalar field potential $V$, the system of equations 
(\ref{finmot}),(\ref{finalP}), (\ref{finalKG}) fully determines the 
evolution of perturbations in matter, 
scalar field and gravitational potentials; it can be numerically 
integrated, given the appropriate initial and boundary conditions. 
It is interesting to note that, for the ``ordinary Quintessence'' models, 
i.e. minimally coupled scalar field, the non-relativistic perturbations 
of the field vanish on sub-horizon scales as a consequence of 
the vanishing anisotropic stress (see Appendix \ref{b} and \ref{c}); 
however, this is generally not true for the extended models we are dealing 
with: the scalar field coupling to the Ricci scalar affects the final 
Poisson equation for the gravitational potentials, and can give rise to 
substantial modifications of the standard structure formation 
picture.  

In the next section, we will consider the special case of non-minimal 
coupling analyzed in \cite{PBM}, adopting the current, local constraints 
on the value of the coupling function $F$.  
 
\section{Extended Quintessence}
\label{specmodel}

Now, let us focus on the ``Extended Quintessence'' model of \cite{PBM}; in 
this model, 
\begin{equation} 
\label{universal}
F = \kappa^{-1} + \zeta \phi^2
\end{equation}  
where $\kappa$ is a universal constant, whose unknown value is 
proportional to the ``bare'' gravitational constant $G^*$; in
scalar-tensor theories, $G^*$ can in principle deviate from the 
Newtonian constant measured in Cavendish-type experiments \cite{EFP}.
Indeed, the actual Newtonian force between two close test masses 
measured at the present time in such experiments is proportional to 
\begin{equation}
\label{Geff}
G_{eff}|_{0}  \equiv  G = {G^* \over \kappa F_0} \left( \omega_{JBD, 0}+2 
\over \omega_{JBD, 0}+3/2 \right) \sim {G^* \over \kappa F_0} \equiv G_N 
\ .
\end{equation}
In the equation above, $\omega_{JBD, 0}$ is the present value of the  
Jordan-Brans-Dicke (JBD) parameter, which reduces to 
\begin{equation}
{1 \over \omega_{JBD}} \equiv  {{F'_0}^2 \over F_0} 
\end{equation}
when the kinetic factor $\omega(\phi)$ in the Lagrangian 
(\ref{Lagr}) is chosen to be a constant, as in our case.  \\
The term between brackets in eq. (\ref{Geff}) is due to the exchange of a 
scalar particle between the two test masses: since the JBD 
parameter is bound from solar-system experiments to be 
larger than $\sim 3000$ (\cite{Will}, \cite{Eub}),  
the present values of  $G_{eff}$ and $G_N$ almost coincide.     
$F$ is the coupling function entering in the Lagrangian of the theory, 
and, as we have seen, it enters in the Friedmann equations and in the 
Poisson equations for the local gravitational potentials;  
the value of this function at present time can be determined only  
{\it locally}, i.e. on the length scales where we are able to    
perform gravitational experiments to determine the ``effective'' 
gravitational constant. Thus, we can give a local representation of the 
function (\ref{universal}): 
\begin{equation}
\label{local}
F  = { 8 \pi G}^{-1}+ \zeta 
(\phi^2-\phi_0^2) \equiv {1 \over 8 \pi G} (1+ y(\phi)) \ ,
\end{equation}
where the dimensionless function $y(\phi) \equiv  8 \pi G \zeta  (\phi^2 
-\phi_0^2) \ll 1$, at low redshifts (see \cite{PB}). Thus,  
on the scales where the effective Newtonian constant reduces to the 
present value of $G$, (\ref{universal}) reduces to (\ref{local}).  \\
In this scheme, the present value of the coupling function reduces 
locally to the inverse of the measured Newtonian constant $F_0= 
(8 \pi G)^{-1}$. \\
We will expand  equations (\ref{finalP}) and 
(\ref{finalKG}) in terms of the perturbative parameter $y(\phi_0)$, where 
the subscript ``0'' refers to the present epoch. Note that 
$$
{1 \over F } = {8 \pi G \over (1+y)} \sim 8 \pi G (1-y) \ \ \ ;
$$ 
for simplicity, we will assume that the scalar field potential 
perturbations $\delta V$ are negligible: in order for this condition to hold, 
even for non-linear scalar field perturbations, the potential must be 
sufficiently flat. If this is the case, eq. (\ref{finalP}) can be written 
as
\begin{equation}
\label{PoissonEQ}
\nabla^2 \Phi = 4 \pi G (1-y) \left[ 
a^2\delta \rho_m 
- {\nabla^2 \delta y \over 8 \pi G} \right] 
\end{equation}

First of all, let us analyze the effect of the background scalar field. 
In the case of an unperturbed non-minimally coupled scalar field, the 
Poisson equation is modified by a 
factor $(1-y)$ with respect to the standard case; this correction is  
scale-independent and proportional to the coupling parameter $\zeta$ at 
any redshift, reducing to zero at the present time. However, at redshifts 
relevant for the onset of structure formation, we expect the gravitational 
force properties to be affected by this correction, which can be positive or 
negative depending on the sign of the coupling parameter $\zeta$.  \\   
In order to appreciate any substantial correction to the standard Poisson 
equation which may be induced by the perturbations in the coupled scalar 
field, we have to evaluate under which conditions, and on which length 
scales $L$, if any, the two terms in square brackets in eq. (\ref{PoissonEQ}) 
are comparable. Note that 
$$
4 \pi G a^2 \delta \rho_m \sim {\cal H}^2 {\delta \rho_m \over \rho_m}
= {L_H}^{-2} \delta_m \ \ ,
$$
$L_H$ being the Hubble length and $\delta_m$ the Dark Matter density 
contrast. 
Now, on the scales of clusters, i.e. length-scales  
$\sim 10^{-3} L_H$, where matter fluctuations 
are becoming non-linear today, in order for the two terms on the rhs of 
eq. (\ref{PoissonEQ}) to be comparable, it would be sufficient to produce 
fluctuations $\delta y \sim 10^{-6}$. On galactic scales, 
overdensities are of order $10^4$, the typical scale being $10^{-5} 
L_H$, so again a perturbation $\delta y \sim 10^{-6}$ would affect 
at a considerable level the Poisson equation. \\
However, we are not free to establish the amount of fluctuations in the 
scalar field component: there are observational constraints, coming from 
the upper limits on the time variation of the gravitational constant and 
from solar-system limits on the JBD parameter 
$\omega_{JBD}$ \cite{Will}.  
The latter is the major constraint, since the Newtonian limit of the 
Klein-Gordon equation is not affected by time variations of the scalar 
field;  we must have 
\begin{equation}
\label{constraint}
{1 \over \omega_{JBD}} \equiv  {{F'_0}^2 \over F_0} = 4 \zeta^2 \phi_0^2
 8 \pi G < 2 \cdot 10^{-4}
\end{equation}
where we selected a conservative lower limit $\omega_{JBD} > 5000 \equiv 
\omega_{lower}$. 
The constraint on $\omega_{JBD}$ is thus translated into an upper 
limit on the combination $\zeta \phi_0$, since $\sqrt{\omega_{JBD}^{-1}} 
= \pm 2 \zeta \phi_0 \sqrt{8 \pi G}$; for positive $\zeta$, it must be 
\begin{equation}
\zeta \phi_0 
\sqrt{8 \pi G} = {1 \over 2 } \sqrt{\omega_{JBD}^{-1}} \lsim 7 \cdot 
10^{-3} = {1\over 2 } 
\sqrt{\omega_{lower}^{-1}}  
\end{equation}
which gives
\begin{equation}
\label{Flimit}
F'_0/ 2 F_0 = {1 \over 2} \sqrt{8 \pi G \omega_{JBD}^{-1}} = \zeta \phi_0 
8 
\pi G
\lsim \sqrt{8 \pi G} \ 7 \cdot 10^{-3} = {\sqrt{8 \pi G 
\ \omega_{lower}^{-1}} \over 2}
\end{equation}   
It is important to emphasize that the condition above has to be thought as 
applying to the total (background plus perturbations) quantities $F, 
F', \phi$ at the present time; i.e. it does not represent a constraint on 
the present value of $F'_{\rm b}/F_{\rm b}$ only, since this value could 
not be observationally discerned from possible local fluctuations;   
rather, the constraint (\ref{Flimit}) applies to the total ratio $F'_0/ 
2 F_0 $, which appears in eq. (\ref{finalKG}).  
As we will see, this reflects into a constraint on $\zeta \phi_0$ 
rather than on $\zeta \phi_{b,0}$. \\
Hereafter, we will omit the subscript ``0'', referring to the 
value at present (or at low-redshifts) of the quantities entering into the 
Klein-Gordon equation. \\  
To check whether solar-system constraints on  $F'/F$ are 
compatible with the requirement $\delta y \sim 10^{-6} $, 
thus allowing for potentially observable imprints on the structure formation
process, we found it convenient to separate the regime of linear 
$\delta \phi \cdot 
\sqrt{8 \pi G} \ll 1$  from the more general case 
$\delta \phi \cdot \sqrt{8 \pi G} \gsim 1 $. 
Let us define the dimensionless scalar field variable 
$\tilde{\phi} \equiv \sqrt{8 \pi G} \phi$. 
In the first case, inserting the upper limit (\ref{Flimit}) into eq. 
(\ref{finalKG}), we have 
\begin{equation}
\label{KGlin}
\nabla^2 \delta \tilde{\phi} \lsim \  {1\over 2 } \sqrt{\omega_{lower}^{-1}}  
\left( 
\delta_m  {\cal H}^2  + 6 \zeta \nabla^2 \delta {\tilde{\phi}} 
\right) \;, 
\end{equation}
where the linearity of $\delta \phi$ allowed to neglect $\delta F$ with 
respect to $F_{\rm b}$ in the factor in front of the brackets, as well as the 
term proportional to $|\delta \phi|^2$; the last 
term on the RHS of eq. (\ref{finalKG}) can be neglected on scales much 
smaller than the horizon. The quantities in eq. (\ref{KGlin}) have to be 
understood as their present values. It is evident that 
\begin{equation} 
\delta {\tilde{\phi}}   \lsim {1\over 2 }
\sqrt{\omega_{lower}^{-1}} \  \delta_m 
\left( {L 
\over L_H} \right)^2  \ \ \ ; 
\end{equation}
as we have seen, one has a typical value of $\delta_m \left({L
/ L_H} \right)^2  \sim 10^{-6}$ on galaxy and cluster scales: 
therefore, the {\it linear} value of $\delta {\phi} \sqrt{8 \pi G}$ on 
those scales is forced to be of the order of $10^{-8}$ (corresponding 
to $\delta y \sim 10^{-11}$), which is orders 
of magnitude below the value required for scalar field perturbations 
to affect the Poisson equation in a considerable way, so that the only effect 
on the Poisson equation would be given 
by the overall factor $(1-y)$, which is active at low redshifts even in the 
absence of field perturbations.  \\ 
To analyze the most general case, we 
will reject the linearity assumption for $\delta \phi$.  
On scales much smaller than the horizon, and neglecting the 
field potential, the Klein-Gordon equation (\ref{finalKG}) for the model 
(\ref{local}) reduces in Fourier space to    
\begin{equation}
\label{KGgen}
 \delta {\tilde{\phi}} = -  { \sqrt{\omega_{JBD}^{-1}}  \over 2} \
\left[  \delta_m \left({L  \over L_H} \right)^2   
+3 \zeta 
\delta {\tilde{\phi}} (2 {\tilde{\phi}} - \delta {\tilde{\phi}} ) +  
\delta {\tilde{\phi}}^2
 \right] \ \ \ ,
\end{equation}
where we used the non-linear expression for  $\delta F  \equiv F-F_{\rm b} = 
\zeta \delta \phi (\delta \phi+2\phi_{\rm b})=\zeta \delta \phi (2 \phi-\delta 
\phi)$.  \\
Without loss of generality, we can assume that 
today $\tilde{\phi} = 1 $,  because we have only a
constraint on the product $\zeta \tilde{\phi}$; with this choice, 
$\zeta < \omega_{lower}^{-1/2} / 2 \sim 10^{-3} $ and eq. 
(\ref{KGgen}) becomes, dividing by $\sqrt{\omega_{JBD}^{-1} } \ne 0 $,   
\begin{equation}
\label{eqsecondog}
 \sqrt{\omega_{JBD} } \  \delta {\tilde{\phi}}  + \delta 
{\tilde{\phi}}^2 
+ \delta_m \left({L  \over L_H}  \right)^2 \sim 0 
\end{equation}
On galaxy and clusters scales, a typical value is  $\delta_m \left({L  
\over L_H}  \right)^2 \sim 10^{-6}$,  much smaller than the lower limit on 
$\sqrt{\omega_{JBD}} \gsim \sqrt{\omega_{lower}} \sim 10^2 $; the two 
solutions of (\ref{eqsecondog}) are therefore 
$ \delta \tilde{\phi} \sim 0 $ (which trivially corresponds to the 
solution previously found under the assumption of linearity), and 
$\delta \tilde{\phi} \sim - \sqrt{\omega_{JBD}} \sim  10^{2}$, which is a
strongly non-linear scalar field perturbation; 
as for the Poisson equation (\ref{PoissonEQ}), the term $\delta y$ is, for 
$\delta {\tilde{\phi}} \sim 10^2$, of order $10^{-1}$, much bigger 
than the value required for the two terms in brackets in eq. 
(\ref{PoissonEQ}) to be comparable. 
The fact that such a value does not depend on the matter overdensity at 
any redshift, makes us argue that it is unphysical, and that the 
Extended Quintessence perturbations can only be linear as long as the 
constraint (\ref{constraint}) applies.  

However, we want to stress that an initially small scalar field 
perturbation could, in this model, grow and become non-linear on 
sub-horizon scales, since the 
limit (\ref{constraint}) is only restricted to the Solar System neighborhood: 
no constraints on spatial fluctuations of the field are available on larger 
scales.  
 
Note again that this kind of ``growing'' solution is only allowed in the 
non-minimally  coupled  case, since we assumed $ \sqrt{\omega_{JBD}^{-1} } 
\ne 0 $: the ``Extended Quintessence'' model could in principle admit 
substantial perturbations of the scalar field, as their time evolution, 
according to eq. (\ref{kleinNR}), is sourced by a non-vanishing term 
which is 
ultimately related to the non-vanishing anisotropic stress 
(see Appendix \ref{b} and \ref{c}) and to the matter perturbations 
themselves.  \\ 
  
\section{Conclusions}
\label{concl}
In this paper we extended the Newtonian approximation to the class of 
scalar-tensor theories of gravity, where a non-minimally coupled scalar 
field is assumed to be responsible for the cosmic acceleration today.  
We obtained the equations relevant for simulations of gravitational 
clustering in these cosmological scenarios. 
The Newtonian Poisson equation acquires new 
contributions from scalar field perturbations, which may turn out to 
affect the gravitational collapse in an interesting way. \\ 
As already argued in a previous study of linear perturbations in these 
models \cite{PB}, a substantial part of scalar field perturbations, along 
with perturbations in its stress-energy tensor components, is powered by 
perturbations in the matter component:  this ``gravitational 
dragging'' is only due to the coupling of the scalar field with the Ricci 
scalar (and, ultimately, with the perturbed matter), and can drive scalar 
field perturbations into the non-linear regime. On the contrary, 
self-interaction originated perturbations in the scalar field will 
unavoidably damp out on the scales relevant for structure formation, as in 
the case of minimally-coupled fields. \\
In order to get specific insights on the role of gravitational dragging, we 
specialized the final equations to the ``Extended Quintessence'' model of 
\cite{PBM}, where the non-minimal coupling strength is quantified by a 
dimensionless coupling parameter.  Without adding a potential for the 
scalar field, we analyzed  the result of the Poisson equation for the 
gravitational potentials and for the scalar field perturbations. 
We isolated and quantified the effect of the time-variation of the 
effective gravitational constant in the equation for clustering 
(\ref{PoissonEQ}), produced by the time evolution of the background value 
of the scalar field.  We also analyzed the effect of perturbations of the 
effective gravitational constant, finding that, in Extended 
Quintessence, the scalar field perturbations are 
prevented from growing non-linear by the Solar System 
constraints on the coupling parameter. Namely, structure formation is 
{\it locally} not affected by the modified gravity at a detectable level;
in principle, however, scalar-field non linearities are not 
precluded even in the Extended Quintessence scenario, at least on those 
scales where the coupling parameter is allowed to vary over a less 
restrictive range of values. Further analysis is required in order to 
quantify the effects on forming structures in more general scalar-tensor 
theories.  
\acknowledgments
F.P. wishes to thank C. Baccigalupi, S.M. Carroll, N. Kaloper and M. White 
for useful hints. During this work, F.P. was supported by a grant of 
Space Science Institute and LBNL of Berkeley, California.   

\appendix
\label{a}
\section{Perturbations in energy density and pressure}
\label{appbackg}
Let us decompose the scalar field stress-energy tensor as follows \cite{PB}:
\begin{equation}  
\nonumber
{T^{i}_{j}}^{(mc)}= \phi^{,i}\phi_{,j}-{1 \over 2}\delta^{i}_{j}
(\phi_{,c}\phi^{,c}+2V)  \ \ ;  \ \
{T^{i}_{j}}^{(nmc)}=F^{,i}_{; j}-\delta^i_j F^{,c}_{;c} \ \ ; \ \
{T^{i}_{j}}^{(grav)}=\left({1 \over \kappa}-F\right) G^{i}_{j} \;.
\nonumber
\end{equation}
The  background and perturbed energy densities and isotropic pressure, 
defined through $T^0_0= -\rho$ and $T^i_i= p/3$, turn out to be
\begin{itemize}  
\item Minimal coupling
\begin{equation}
\nonumber
{\rho_{\phi}}_{\rm b}^{(mc)}= {1 \over 2 a^2} 
\dot{\phi}_{\rm b}^2 + V(\phi_{\rm b}) \ \ ; \ \
{p_{\phi}}_{\rm b}^{(mc)}={1 \over 2 a^2} 
\dot{\phi}_{\rm b}^2 - V(\phi_{\rm b}) \ \ ; 
\nonumber
\end{equation}
\begin{equation}
\nonumber
\delta \rho_{\phi }^{(mc)}=  {1 \over 2 a^2}  (1-2\Phi)
({\delta \dot{\phi}}^2 +2 \dot{\phi}_{\rm b}{\delta \dot{\phi}}) +
 {1 \over 2 a^2}  (1+2\Psi) |\nabla \delta \phi|^2+ \delta V
- \Phi {{\dot{\phi_{\rm b}}}^2 \over a^2} 
\nonumber
\end{equation}
\begin{equation}
\nonumber
\delta p_{\phi}^{(mc)}={1 \over 2 a^2}  (1-2\Phi)({\delta 
\dot{\phi}}^2 +2 \dot{\phi}_{\rm b}{\delta \dot{\phi}})- {1 \over 6 a^2}  
(1+2\Psi) 
|\nabla \delta \phi|^2 -\delta V- \Phi{ {\dot{\phi_{\rm b}}}^2 \over  a^2}
\nonumber
\end{equation}
\item Non-minimal coupling
\begin{equation}
\nonumber
{\rho_{\phi}}_{\rm b}^{(nmc)}= 
-3{{\cal H} \over a^2}\dot{F}_{\rm b} \ \ ; 
\ \
{p_{\phi}}_{\rm b}^{(nmc)}={1 \over  a^2} \ddot{F}_{\rm b} + 
{{\cal H} \over a^2} 
\dot{F}_{\rm b} \ \ ; 
\nonumber
\end{equation}
\begin{equation}
\nonumber
\delta \rho_{\phi }^{(nmc)}=  {1 \over 2 a^2} 
[3 \dot{F}_{\rm b} ( \dot\Psi+2 {\cal H} \Phi ) +3 \delta 
\dot{F} ( \dot\Psi +2 {\cal H} \Phi -{\cal H})
-\nabla \Psi\nabla\delta F
 +(1+2 \Psi)\nabla^2 \delta  F ]
\nonumber
\end{equation}
\begin{equation}
\nonumber
\delta p_{\phi}^{(nmc)}=
-{2 \over a^2} \Phi \ddot{F}_{\rm b}+{1 \over a^2}(1-2 \Phi)\delta 
\ddot{F} -{ {\dot{F}}_{\rm b} \over a^2} ({\cal H} 
-2{\cal H} \Phi-\dot\Phi - 2\dot \Psi)+{1 \over 3 a^2}
\nabla\delta F\nabla(2\Psi -3\Phi)
-{2 \over 3 a^2}(1+2 \Psi) \nabla^2 \delta F
\nonumber
\end{equation}
\item Gravitational 
\begin{equation}
\nonumber
{\rho_{\phi}}_{\rm b}^{(grav)}=  
3{{\cal H}^2 \over a^2}(\kappa^{-1}-{F}_{\rm b}) \ \ ; 
\ \
{p_{\phi}}_{\rm b}^{(nmc)}=-{1 \over  a^2} (\kappa^{-1}-{F}_{\rm b}) 
(2 \dot{\cal H} +{\cal H}^2) \ \ ; 
\nonumber
\end{equation}
\begin{equation}
\nonumber
\delta \rho_{\phi }^{(grav)}= 
-3{ {\cal H}^2 \over a^2} \delta F+
{1 \over a^2}(\kappa^{-1}-{F}_{\rm b} - \delta F)(2 \nabla^2\Psi -6 
{\cal H}^2 \Phi -6 {\cal H} \dot\Psi)
\nonumber
\end{equation}
\begin{equation}
\nonumber
\delta p_{\phi}^{(grav)}=
(2 \dot{\cal H}+{\cal H}^2) {\delta F \over 
a^2}+{2 \over a^2}(\kappa^{-1}-{F}_{\rm b} - \delta F) \left[ (2 
\dot{\cal H}+{\cal H}^2)\Phi + {\cal H}\dot\Phi
+\ddot{\Psi} +2 {\cal H} \dot{\Psi}+{1 \over 3 } 
\nabla^2(\Phi - \Psi) \right]
\nonumber
\end{equation}
\end{itemize}
In the equations above, the subscript $b$ refers to background 
quantities. \\
In the Newtonian limit, we have 
\begin{equation}
\label{nlmc}
(\delta \rho_{\phi}+ 3 \delta p_{\phi})^{(mc)} \rightarrow  -2 \delta V
\end{equation}
\begin{equation}
\label{nlnmc}
(\delta \rho_{\phi}+ 3 \delta p_{\phi})^{(nmc)} \rightarrow  - {\nabla^2 
\delta F \over a^2}
\end{equation}
\begin{equation}
\label{nlgrav}
(\delta \rho_{\phi}+ 3 \delta p_{\phi})^{(grav)} \rightarrow  
 (\kappa^{-1}-F){2 \nabla^2 \Phi \over a^2} \ \ \ .
\end{equation}
At the same time, it is easy to verify that the Newtonian limit of the 
perturbed Einstein equation $\delta R^0_0 = \kappa \delta S^0_0$ gives 
\begin{equation}
\label{nlpoisson}
\nabla^2 \Phi=  {a^2 \over 2} \kappa (\delta \rho+3 \delta p)
\end{equation}
where $\delta \rho$ and $\delta p$ include contributions from matter and 
from scalar field (minimal coupling, non-minimal coupling and 
gravitational parts). Substituting the limits (\ref{nlmc})-(\ref{nlgrav})
into (\ref{nlpoisson}), and comparing with eq. (\ref{finalP}) we see that 
the contribution $-2a^2 \delta V$ in eq. (\ref{finalP}) originates from 
the minimal coupling, the term  $-\nabla^2 \delta F$ originates from the 
non-minimal coupling, while the  gravitational contribution can be 
absorbed in the quantity $F^{-1}$. Looking at eq. (\ref{finalP}), we see 
that the role of the non-minimal coupling is twofold: on the one hand, it 
generates the energy and pressure contribution $-\nabla^2 \delta F$, 
which is non-vanishing only if the scalar field 
fluctuations are non vanishing; on the other hand, it also generates a 
modification of the gravitational constant, where the effective 
gravitational constant $F^{-1}$ is generally different from $\kappa$ even 
if the scalar field does not have fluctuations. 

\section{Anisotropic stress}
\label{b}
The unperturbed space-space component of the Einstein tensor has only 
trace component: 
\begin{equation}
\label{gijb}
{G^i_j}_{\rm b}= -{2 \over a^2} (\dot{\cal H}+{1 \over 2} {\cal H}^2) 
\delta^i_j
\end{equation}
On the other hand, the perturbations $\delta {G^i_j}$ have a trace and a 
traceless part: 
\begin{equation}
\label{deltagijtrace}
\delta {G^i_j}_{(trace)}= {2 \over a^2} \delta^i_j 
\left[ (2 \dot{\cal H} +{\cal H}^2 ) \Phi
+{\cal H}\dot \Phi + \ddot\Psi+2 {\cal H} \dot\Psi
+{1 \over 3}(\nabla^2 \Phi- \nabla^2 \Psi) \right] 
\end{equation}
and
\begin{equation}
\label{deltagijtraceless}
\delta {G^i_j}_{(traceless)}={1 \over a^2}\left[
-\gamma^{ik}(\Phi - \Psi)_{,jk}+{1 \over 3 } \nabla^2
(\Phi - \Psi) \delta^i_j \right] +(vect. \ and \   tensor 
\ pert.  )
\end{equation}
Here, we allowed for tensor and vector perturbations, which we do not 
write explicitly (for details, see \cite{KS}).

The total stress energy tensor for the scalar field in the Extended 
Quintessence model can be decomposed as 
$$
T^{\mu}_{\nu}={T^{\mu}_{\nu}}^{(mc)}+{T^{\mu}_{\nu}}^{(nmc)}+
{T^{\mu}_{\nu}}^{(grav)}
$$

The anisotropic stress is defined in terms of the
spatial components of the stress-energy tensor; the latter can always be 
decomposed as the sum of a trace tensor and a traceless tensor: 
\begin{equation}
\label{tij}
T^i_j \equiv p_{\rm b} [ \delta^i_j+ \pi_L \delta^i_j + 
{\pi_T}^i_j] \;, 
\end{equation}
where $p_{\rm b}$ is the fluid pressure in the unperturbed state; 
$\pi_L$ is the isotropic pressure perturbation (a commonly used 
definition is $\pi_L \equiv \delta p / p_{\rm b} $), 
and $p_{\rm b}{\pi_T}^i_j$ is the 
anisotropic stress tensor, defined as the traceless part of $T^i_j$: 
$$
p_{\rm b}{\pi_T}^i_j \equiv T^i_j- {1 \over 3 } \delta^i_j T^k_k \ \ .
$$  
We are working in configuration space, so $\pi_L$ and ${\pi_T}^i_j$ 
are functions of $({\bf x}, \eta)$. \\ 
Note that in linear theory we can write
$$
{\pi_T}^i_j=  (\gamma^{ik} \partial_i \partial_k -{ \delta^i_j  \over 3 } 
\nabla^2 ) \pi_T
$$
$\pi_T$ being the amplitude of the anisotropic stress perturbation, 
see eg. 
\cite{MB}, \cite{KS}). \\

In the following, we will consider linear perturbations 
in the metric (described in the conformal Newtonian gauge), but the 
scalar field  perturbations will generally be non linear.  \\
For the minimally coupled field, with $\omega=1$, the tensor 
${T^{i}_{j}}^{(mc)}$ is given 
by
$$
{T^{i}_{j}}^{(mc)}= \phi^{,i}\phi_{,j}-{1 \over 2}\delta^{i}_{j}
(\phi_{,c}\phi^{,c}+2V)= {T^{i}_{j}}^{(mc)}_{(trace)}
+{T^{i}_{j}}^{(mc)}_{(traceless)}
$$
where
$$ 
{T^{i}_{j}}^{(mc)}_{(trace)}= 
{1 \over 3 a^2}(1+2 \Psi)|\nabla \delta \phi|^2 \delta^i_j-
{1\over 2} \delta^i_j (\phi_{,c}\phi^{,c}+2V)
$$
and
$$
{T^{i}_{j}}^{(mc)}_{(traceless)}= 
{1 \over  a^2}(1+2 \Psi) \gamma^{i k }\partial_k \delta \phi 
\partial_j \delta \phi-{1 \over 3 a^2}(1+2 \Psi)|\nabla 
\delta \phi|^2 \delta^i_j
$$
Note that, since $\phi_{\rm b}$ does not depend on the spatial coordinates,  
$\partial_i \phi = \partial_i \delta \phi$; for this reason, in the 
linear regime (i.e., for linear perturbations of the scalar field), the 
anisotropic stress from a  minimally coupled scalar field is negligible, 
being second order in the scalar field perturbations. In 
general, the anisotropic stress from a minimally coupled scalar field is 
of order ${\cal{O}}(|\nabla \delta \phi|)^2$.

For the non-minimal coupling tensor, 
$$
{T^{i}_{j}}^{(nmc)}=F^{,i}_{; j}-\delta^i_j F^{,c}_{;c}= 
{T^{i}_{j}}^{(nmc)}_{(trace)}+
{T^{i}_{j}}^{(nmc)}_{(traceless)} 
$$
with 
$$
{T^{i}_{j}}^{(nmc)}_{(trace)}=
-{1\over 3 a^2}\delta^i_j \left[\nabla\delta F 
\nabla\Psi - (1+2\Psi)  \nabla^2 \delta F + \dot{F}(3 {\cal H} -3 
\dot\Psi
-6 {\cal H}\Phi )\right] - \delta^i_j {F^{,c}}_{;c}
$$
and, after some algebra, it can be shown that 
\begin{eqnarray}
\nonumber
{T^{i}_{j}}^{(nmc)}_{(traceless)}&=&  {1 \over a^2} (1+2\Psi) 
\left[\gamma^{ik}F_{,kj}-{1\over 3}
 \delta^i_j \nabla^2 F \right]+
{1 \over a^2}\gamma^{ik}\Psi_{,j}F_{,k}+
\gamma^{i \lambda} \Psi_{, \lambda} \gamma_{kj} F^{,k}-
{2 \over 3 a^2}\delta^i_j\nabla F \nabla\Psi = \\
\nonumber
&=&{(1+2 \Psi) \over a^2}\left[\gamma^{ik}\delta F_{,kj}-{1\over 
3}\delta^i_j \nabla^2 \delta F  \right]
+{2 \over a^2} \left[\gamma^{ik} \Psi_{,j} \delta F_{,k}
-{1 \over 3 }\delta^i_j \nabla\delta F \nabla\Psi  
\right]
\nonumber
\end{eqnarray}
We see from here that the anisotropic stress from the non-minimally 
coupled field contains terms which are proportional to the spatial 
gradients of $F$; in general, these terms are of the order 
${\cal{O}}(|\nabla \delta \phi|)$, so the anisotropic stress for a 
non-minimally coupled  scalar field would survive even in the case of 
linear perturbations of the field. 

Finally, the gravitational term is 
$$
{T^{i}_{j}}^{(grav)}=\left({1 \over \kappa}-F\right) G^{i}_{j}=
{T^{i}_{j}}^{(grav)}_{(trace)} + {T^{i}_{j}}^{(grav)}_{(traceless)}
$$
Using equations (\ref{gijb}), (\ref{deltagijtrace}) and 
(\ref{deltagijtraceless}), we have 
$$
{T^{i}_{j}}^{(grav)}_{(trace)}= \left({1 \over \kappa}-F\right)
{2 \over a^2}\delta^i_j
\left[ - (\dot{\cal H}+{1 \over 2} {\cal H}^2)+
(2 \dot{\cal H} +{\cal H}^2 ) \Phi
+{\cal H}\dot\Phi + \ddot\Psi + 2 {\cal H} \dot\Psi
+{1 \over 3}(\nabla^2 \Phi- \nabla^2 \Psi) \right]
$$
and
$$
{T^{i}_{j}}^{(grav)}_{(traceless)}= 
\left({1 \over \kappa}-F\right)
{1 \over a^2}\left[
-\gamma^{ik}(\Phi - \Psi)_{,jk}+{1 \over 3 } \nabla^2
(\Phi - \Psi) \delta^i_j \right] +(vect. \ and \    tensor
\ pert.  )
$$
Therefore, the total anisotropic stress, without taking into account 
vector and tensor perturbations of the metric, will be given by 
the sum of the three contributions from the minimal coupling, the 
non-minimal coupling and the gravitational terms. Since its  
background value is zero, we have 
\begin{eqnarray}
\nonumber
\delta {T^i_j}_{traceless}^{mc+nmc+grav} &=& 
{1 \over  a^2}(1+2 \Psi) \gamma^{i k }\partial_k \delta \phi 
\partial_j 
\delta \phi-{1 \over 3 a^2}(1+2 \Psi)|\nabla \delta \phi|^2 
\delta^i_j + \\
\nonumber 
&+&
{(1+2 \Psi) \over a^2}\left[\gamma^{ik}\delta F_{,kj}-{1\over 
3}\delta^i_j \nabla^2 \delta F  \right]
+{2 \over a^2} \left[\gamma^{ik} \Psi_{,j} \delta F_{,k}
-{1 \over 3 }\delta^i_j \nabla\delta F \nabla\Psi  
\right] + \\
\nonumber
&+& \left({1 \over \kappa}-F\right)
{1 \over a^2}\left[
-\gamma^{ik}(\Phi - \Psi)_{,jk}+{1 \over 3 } \nabla^2
(\Phi - \Psi) \delta^i_j \right] +(vect. \ and \    tensor
\ pert.  )
\end{eqnarray}
Using eq. (\ref{deltagijtraceless}), and writing the gravitational 
contribution on the left-hand side of the traceless
part of the perturbed space-space Einstein equation
$$
{\delta G^i_j}_{(traceless)}= \kappa \ \delta {T^i_j}_{(traceless)}
$$
we obtain:  
\begin{eqnarray}
& F & \left[-\gamma^{ik}(\Phi - \Psi)_{,jk}+{1 \over 3 } 
\nabla^2
(\Phi - \Psi) \delta^i_j \right] = 
(1+2 \Psi) \gamma^{i k }\partial_k \delta \phi \partial_j 
\delta \phi
-{1 \over 3 a^2}(1+2 \Psi)|\nabla \delta \phi|^2 \delta^i_j + \\
\nonumber 
&+& {(1+2 \Psi) \over a^2}\left[\gamma^{ik}\delta F_{,kj}-{1\over 
3}\delta^i_j \nabla^2 \delta F  \right]
+{2 \over a^2} \left[\gamma^{ik} \Psi_{,j} \delta F_{,k}
-{1 \over 3 }\delta^i_j \nabla\delta F \nabla\Psi  
\right]  \ \ , 
\end{eqnarray}
where contributions from vectors and tensors have been omitted. \\
As an important consequence, we can note that, in the case of a minimally 
coupled scalar field (i.e., $F \equiv \kappa^{-1}$), one has
$$
\Phi= \Psi + {\cal{O}}( \delta \phi)^2 \ \ ;
$$
if $\delta \phi$ are linear perturbations in the minimally
coupled scalar field, the difference between the two gravitational 
potentials is negligible in a first-order theory.    

\section{Newtonian approximation for the minimally coupled scalar field 
in linear perturbation theory}
\label{c}
In order to better understand the important role of the non-minimal 
coupling, let us see what to expect to be the perturbation behavior in 
the case of a minimally coupled scalar field.  \\
First of all, in the minimally coupled case, the anisotropic pressure 
perturbation  is of order  ${\cal{O}}( \delta \phi)^2$, constraining 
the two gravitational  potentials $\Phi$ and $\Psi$ to differ by 
terms of the same order.  Thus, we expect that the (Newtonian) Poisson 
equations  (\ref{finalPsi}), (\ref{finalP}) will be equivalent in linear 
theory, for a minimally coupled scalar field. \\
Taking the trace of the perturbed space-space Einstein equation 
$ \delta G^i_i = \kappa \delta T^i_i $
we obtain, for the case of minimally-coupled scalar field ($\kappa 
\equiv 8 \pi G_N) $, 
\begin{equation}
\label{einsteinij}
 (2 \dot{\cal H} +{\cal H}^2 ) \Phi
+{\cal H}\dot \Phi+ \ddot\Psi + 2 {\cal H} \dot\Psi
+{1 \over 3} (\nabla^2 \Phi - \nabla^2 \Psi) 
= {\kappa  \over 2} \left[ {1 \over 2} (1-2 \Phi)
(\delta \dot{\phi}^2 + 2\dot \phi_{\rm b} \delta \dot{\phi}) 
- \Phi\dot \phi_{\rm b}^2 -{1 \over 6} |\nabla\delta \phi|^2
-a^2\delta V \right] \ \ \ .
\end{equation}
Here  $\nabla^2 (\Phi - \Psi) 
\sim {\cal{O}}{|\nabla \delta \phi|^2}$ can be neglected for linear 
perturbation of the field. As a consequence,  $a^2 \delta V$ will be of 
order ${\cal H}^2$ (in eq. (\ref{einsteinij}), $\delta \phi$ denotes 
non-relativistic perturbations); by differentiating 
$a^2 \delta V$ with respect to the conformal time, one can show that 
$a^2 \delta V' \sim {\cal{O}}({\cal H}^2)$: inserting this result in 
the Newtonian limit of the perturbed Klein-Gordon equation for the 
non-relativistic scalar field perturbations (\ref{kleinNR}), we can see 
that, for linear perturbations $\delta \phi$,    
$$
\nabla^2 \delta \phi  \sim a^2 \delta V' \sim   
{\cal{O}}({{\cal H}}^2) \ \ ,
$$
which is negligible in the Newtonian limit of scales smaller than the 
horizon ; thus, the Klein-Gordon equations implies that the (minimally 
coupled) scalar field perturbations, in linear theory, will be negligible 
on those scales. For this reason, the scalar field will behave as a  
homogeneous component on the scales relevant for structure formation. 
Correspondingly, $\Phi  \rightarrow \Psi$, and the two 
equations (\ref{finalP}) and (\ref{finalPsi}), with $F=(8 \pi G^*)^{-1}$, 
will be identical (up to terms of order (${\cal H}^2, \delta \phi^2$)).


\end{document}